\documentclass[useAMS,usenatbib]{mn2e}

\usepackage{lastpage}
\usepackage{times}
\usepackage{amsmath,amsfonts,graphicx}
\usepackage{color}

\newcommand{\msun}{\mbox{M}_\odot}
\newcommand{\rsun}{\mbox{R}_\odot}

\renewcommand{\d}{\mbox{d}}
\newcommand{\s}{\mbox{s}}

\newcommand{\kic}{KIC~2856960} 
\newcommand{\kepler}{\emph{Kepler}}

\title[\kic: the impossible triple star]{\kic: the impossible triple star%
}
\author[Marsh, Armstrong \& Carter]{T.\ R.\ Marsh$^{1}$, D.\ J.\ Armstrong$^{1}$, 
P.J.\ Carter$^{2,1}$\\
$^1$Department of Physics, University of Warwick,
Gibbet Hill Road, Coventry, CV4 7AL, UK\\
$^2$School of Physics, H.H.\ Wills Physics Laboratory,
University of Bristol, Tyndall Avenue, Bristol BS8~1TL
}
\begin{document}

\date{Accepted ----. Received ----; in original form ----}

\pagerange{\pageref{firstpage}--\pageref{lastpage}} \pubyear{2014}

\maketitle

\label{firstpage}

\begin{abstract}
  \kic\ is a star in the \kepler\ field which was observed by \kepler\ for 4
  years. It shows the primary and secondary eclipses of a close binary of
  period $0.258\,\d$ as well as complex dipping events that last for about
  $1.5\,\d$ at a time and recur on a $204\,\d$ period. The dips are thought to
  result when the close binary passes across the face of a third star. In this
  paper we present an attempt to model the dips. Despite the apparent
  simplicity of the system and strenuous efforts to find a solution, we find
  that we cannot match the dips with a triple star while satisfying Kepler's
  laws. The problem is that to match the dips the separation of the close
  binary has to be larger than possible relative to the outer orbit given the
  orbital periods. Quadruple star models can get round this problem but
  require the addition of a so-far undetected intermediate period of order $5$
  -- $20\,\d$ that has be a near-perfect integer divisor of the outer $204\,\d$
  period. Although we have no good explanation for \kic, using the full set of
  \kepler\ data we are able to update several of its parameters. We also
  present a spectrum showing that \kic\ is dominated by light from a K3- or
  K4-type star.
\end{abstract}


\section{Introduction}
A large fraction of stars are found in binary systems, and a significant
number of binary stars reside in triple systems.  Triple stars add complexity
to the dynamics and evolution of stars
\citep{2013MNRAS.431.2155N,2001ApJ...562.1012E}. Even though binary stars
offer many outcomes closed to single star evolution, there are systems where
evolution within a triple is the simplest explanation for otherwise puzzling
data \citep{2001ApJ...563..971O}, and location within triple systems has been
suggested as a way to speed the merger of compact objects, which might help
drive Type~Ia supernovae and other exotic transients
\citep{2011ApJ...741...82T}.  Triple stars are mini-clusters, with three
co-eval stars in orbits which, in favourable circumstances, may allow us to
determine precision fundamental parameters for all three objects.

Eclipsing systems are a well-travelled route to precision stellar
parameters. In the case of triple systems there are three different pairs of
stars that can eclipse, but the chances of suitably aligned systems are low,
given the hierarchical structure of triples which contain binary stars in much
longer period and therefore wider orbits with third stars. Fortunately, the
nearly uninterrupted coverage provided by the \kepler\ satellite has uncovered
a significant number of triples
\citep{2012AJ....143..137G,2013ApJ...768...33R,2014AJ....147...45C}, and a
number of these are multiply eclipsing. Examples are KOI-126, which has a
$1.77\,\d$ close binary in a $33.9\,\d$ orbit with a third star
\citep{2011Sci...331..562C}, and HD~181068, which contains a $0.90\,\d$ binary
in a $45\,\d$ orbit with a red giant \citep{2011Sci...332..216D}. KOI-126 in
particular led to precise masses and radii of all three component stars.

\kic\ is another eclipsing triple star observed by \kepler. Listed as an
eclipsing binary by \cite{2011AJ....141...83P}, \kic\ was subsequently found
to be a triple system after the discovery of a second set of eclipses in
addition to those of the binary \citep{2012A&A...545L...4A}.  The binary in
\kic\ reveals itself through $\sim 1$\%-deep eclipses (primary and secondary)
on a period of $0.258\,\d$. Its triple nature is apparent from complex
clusters of dips in flux up to $8$\%\ deep, which last for a little over one
day at each appearance, and recur on a period of $\sim 204\,\d$
\citep{2012A&A...545L...4A,2013ApJ...763...74L}.  \cite{2012A&A...545L...4A}
suggested two very different models for the system. In the first, the dip
clusters are produced when a dim circumbinary object, possibly a planet,
passes in front of the close binary, with multiple eclipses taking place as
the binary completes its orbits. In the second, it is the close binary passing
across the face of a third star that produces the dips. The second model was
proven correct by \cite{2013ApJ...763...74L} who found variations in the times
of the eclipses of the close binary consistent with light travel time
variations as it orbited a third star. Thus \kic\ has all the characteristics
of a hierarchical triple, with a close binary of a period $\sim 0.258\,\d$ in
a $204\,\d$ period orbit with a third star. 

\begin{figure*}
\hspace*{\fill}
\centering
\includegraphics[angle=270,width=\textwidth]{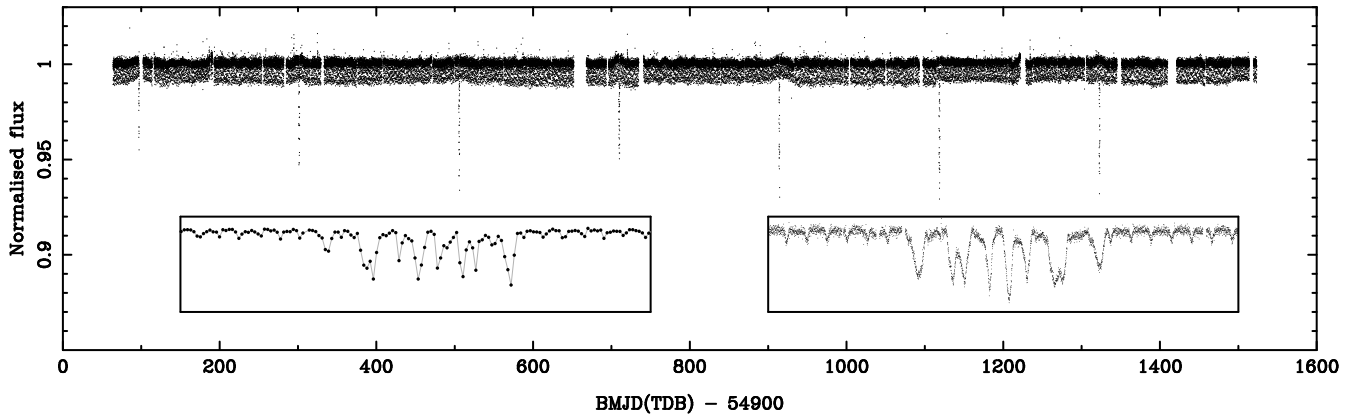}
\hspace*{\fill}

\caption{The light-curve of \kic\ observed by \kepler. This spans 4
years and was mostly taken in long cadence (30 min sampling) mode. The main
section of short cadence (1 min) data runs from 1100 to $1300\,\d$ on the plot,
with an additional short section at the end. For clarity, the data were binned
in a way which does not affect the long cadence data, but placed the short
cadence data onto a similar sampling (50 points per day. The seven narrow
dropouts are the ``dips''. The insets show 3 day-long zoomed plots around the
third set of dips centred at $505\,\d$ (left) and the sixth set of dips at 
$1119\,\d$ (right), the only one observed in short cadence. The latter zoomed
section is displayed at the full short-cadence resolution. Each zoom runs
vertically from 0.90 to 1.02. In the left-hand zoom, we join the dots to
make the dips clearer. Data away from the dips has a bi-modal
appearance caused by the binary eclipses which can also be seen in the
insets.
\label{fig:overview}}
\end{figure*}

There has been no analysis of \kic\ to see if, like KOI-126, it can yield
precision parameters of its component stars. Here we document our efforts to
do this, efforts which ended in failure. Not only do we not find a precision
set of masses and radii for the component stars, we do not find \emph{any}
physically-consistent set of masses and radii that comes close to explaining
\kic's light curve. The nature of the disagreement leads us to conclude that,
despite appearances, \kic\ cannot be modelled as a triple star. Here we describe
why we are led to this conclusion, and tentatively propose quadrupole star
models as a possible escape route.

\section{Observations}
The NASA \emph{Kepler} satellite is a mission producing extremely high
precision, near continuous light curves of $\sim$155,000 stars on the level
of 20 ppm \cite{2010ApJ...713L..79K,2010ApJ...713L.109B,2010ApJ...713L..79K}
The mission began science operations on 13 May 2009.

\kepler\ added significant new data on \kic\ to those published by
\cite{2012A&A...545L...4A,2013ApJ...763...74L}. Particularly significant are
somewhat more than two quarters of short cadence (1 minute) data which fully
resolve both the binary eclipses and one set of dips, compared to the majority
of the data which were taken in long cadence mode (30 minutes). Sadly,
\kepler\ suffered a failure of one its reaction wheels shortly before a second
set of dips were to be observed in short cadence, terminating its coverage of
\kic. This left ~1500\ d of data publicly available on the NASA Data
Archive\footnote{http://archive.stsci.edu/kepler/}, from which we sourced the
light curve of \kic. Detrending of the data was performed by the Kepler
science team using the Pre-search Data Conditioning pipeline (PDC-MAP, see
\cite{2012PASP..124..985S} for an overview and examples, and
\cite{2012PASP..124.1000S} for a full description of the detrending process).

We obtained spectra of \kic\ in service mode on the night of May 21, 2012.
The spectra were acquired with the ISIS spectrograph on the 4.2m William
Herschel Telescope at the Roque de Los Muchachos on the island of La~Palma in
the Canary Islands. ISIS uses a dichroic and two separate arms to cover
blue and red wavelengths simultaneously. We used the 600 lines/mm gratings to
cover the wavelength ranges 370 to $530\,$nm and $565$ to $735\,$nm at
resolutions of $0.20$ and $0.18\,$nm respectively, with around 4 pixels per
resolution element. We took two spectra in each arm with $1200\,\s$
exposures, separated by 3 hours in time. There was no difference or radial
velocity shift between the spectra so we combined them into one with $2400\,\s$
total exposure in each arm.

\section{Analysis}

\subsection{An overview of the light-curve of \kic}
Fig.~\ref{fig:overview} 
displays an overview of the \kepler\ light-curve of \kic. For most of the time,
the light-curve displays $\sim 1$\% deep eclipses which repeat every
$0.129\,\d$ which are the primary and secondary eclipses of a
$P = 0.258\,\d$ binary star. \kic\ would be unremarkable were it not for
seven brief intervals during which the flux dips up to 8\%\ below its
normal level. These are the ``dips'', previously referred to, which recur every
$204\,\d$. As the insets of two of these dips show, they have a complex
structure in which the flux sometimes returns to its normal level between dips
and each cluster contains up to 9 minima. Our aim is to try to elucidate how
these structures come about.

\subsection{The close binary and its orbit within the triple}
\label{sec:binary}
We begin our analysis by looking at the close binary light curve and the light
travel time variations. These are the most secure aspect of the system, and
given the difficulties we encounter understanding the dips, it is desirable in
the first instance to understand as much a possible about the light-curve away
from the dips. This is a repeat of the work of \cite{2013ApJ...763...74L} but
with the considerable advantage of the short cadence data which allows us to
resolve the binary eclipses. Before starting, we first define all the
parameters that define the triple star model (Table~\ref{tab:param}). We adopt
the convention that the $P = 0.258\,$d close binary and its orbit will be
referred to as ``the binary'', while the $P = 204\,$d long period outer orbit
will be called ``the triple''.
\begin{table*}

\begin{tabular}{llll}
Name & Unit & Description & Comment\\
\hline
$R_1$ & $\rsun$ & Radius of star 1 of the binary&\\
$R_2$ & $\rsun$ & Radius of star 2 of the binary&\\
$R_3$ & $\rsun$ & Radius of star 3, the tertiary component of the system&\\
$a_1$ & $\rsun$ & Semi-major axis of star 1 within the binary&\\
$a_2$ & $\rsun$ & Semi-major axis of star 2 within the binary&\\
$a_3$ & $\rsun$ & Semi-major axis of star 3 within the triple&\\
$a_b$ & $\rsun$ & Semi-major axis of binary within the triple&\\
$i_b$ & degrees & Orbital inclination of the binary&\\
$i_t$ & degrees & Orbital inclination of the triple&\\
$T_b$ & days    & Epoch of primary eclipse of close binary, star~1 
transiting star 2, MJD(BTDB)&\\
$T_t$ & days & Epoch of the dips, close binary transiting star 3, MJD(BTDB)&\\
$P_b$ & days & Orbital period of the binary&\\
$P_t$ & days & Orbital period of the triple&\\
$\Omega_b$ & degrees & Longitude of ascending node of the binary&\\
$\Omega_t$ & degrees & Longitude of ascending node of the triple& 
Fixed to $270^\circ$\\
$e_b$ & --- & eccentricity of the binary & Fixed to $0$\\
$e_t$ & --- & eccentricity of the triple & \\
$\omega_b$ & degrees & Longitude of periastron of the binary &
Fixed to $0^\circ$\\
$\omega_t$ & degrees & Longitude of periastron of the triple&\\
$S_1$ & $\rsun^{-2}$ & Central surface brightness, star 1&\\
$S_2$ & $\rsun^{-2}$ & Central surface brightness, star 2&\\
$S_3$ & $\rsun^{-2}$ & Central surface brightness, star 3&\\
$u_1$ & --- & Linear limb darkening coefficient, star 1&Fixed to $0.5$\\
$u_2$ & --- & Linear limb darkening coefficient, star 2&Fixed to $0.5$\\
$u_3$ & --- & Linear limb darkening coefficient, star 3&Fixed to $0.5$\\
$l_3$ & --- & ``Third light'' as a fraction of total flux&\\
\hline
\end{tabular}
\caption{Physical parameters defining the triple star
  model. \label{tab:param}}
\end{table*}
We will exclusively use the symbols defined in Table~\ref{tab:param}. These
differ in some respects from those defined by \cite{2013ApJ...763...74L} and
care should be taken when comparing our values to theirs. When we quote their
values, we have translated the symbols they used into our convention. We
define the ``ascending node'' to be the point in an orbit when a star lies in
the plane of the sky that contains the focus of its orbit and is travelling
away from the observer.

\begin{figure}
\hspace*{\fill}
\centering
\includegraphics[angle=270,width=\columnwidth]{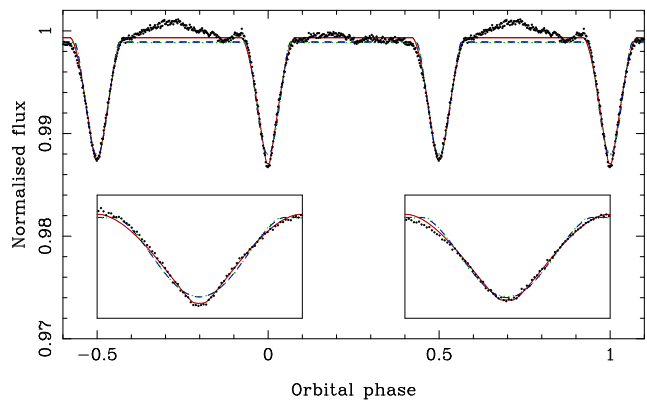}
\hspace*{\fill}

\caption{The short cadence data of \kic\ (291505 points) folded on the 
$0.258\,$d period of the
close binary, excluding data affected by the dips, and corrected for
light-travel time variations induced by its orbit within the triple system.
The data have been averaged into 500 bins per orbit.
Phase~0 is defined by the slightly deeper primary eclipse. While there are
some signs of tidally-induced ellipsoidal modulations outside the eclipses,
these are largely masked by what are probably starspot-induced variations. 
Insets show horizontally-expanded views of the primary (left) and secondary
(right) eclipses. The solid line shows a best-fitting model based upon two
limb-darkened spheres. This has a radius ratio $R_1/R_2 = 0.88$. The dashed
and dotted lines shows models of the same total eclipse width but with the
radius ratio held fixed at $R_1/R_2 = 2$ and $R_1/R_2 = 4$ respectively.
\label{fig:binary-fold}}
\end{figure}

\begin{figure}
\hspace*{\fill}
\centering
\includegraphics[angle=270,width=\columnwidth]{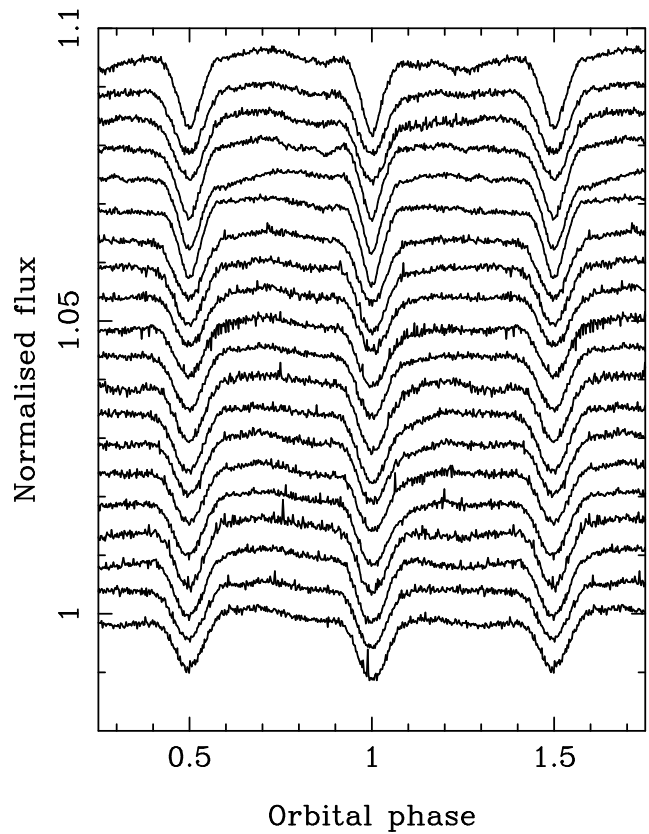}
\hspace*{\fill}

\caption{The close binary light curve folded in 20 time intervals spread
equally from start to end of the \kepler\ observations of \kic. Time increases
upwards, The topmost light curve and those 3 to 5 places down from it are
derived from short-cadence data. There are obvious variations of shape outside
the eclipse, e.g. the light curve fifth from the top. The light curves are
offset vertically by $0.005$.
\label{fig:multi-binary-fold}}
\end{figure}

\cite{2013ApJ...763...74L} used the first 6 quarters of \kepler\ data on \kic,
all taken in long cadence. They modelled the binary light curve with two
tidally-distorted stars in Roche geometry, finding that it matches a pair of
low-mass M dwarfs, but with a $97$\% ``third light'' contribution, which they
ascribed to the third star. Importantly, they found that the eclipse times of
the binary exhibited a large periodic variation, on the same period ($204\,$d)
as the dips.  This variation can only be explained through light travel time
variations. The orbit defined by the light travel time variations was one of
high eccentricity with $e_t = 0.612 \pm 0.082$, and a periastron angle
indicating that the major-axis of the ellipse lies close to the plane of the
sky.  Fig.~\ref{fig:binary-fold}
shows the phase-folded light curve of the short cadence data taken towards the
end of \kepler's coverage of \kic. The variations outside eclipse are
suggestive of star spots. Fig.~\ref{fig:multi-binary-fold}
shows phase-folded light curves from data in 20 time intervals from the start to the end of the 4 years of \kepler\
data. The sharper short-cadence dominated curves are visible towards the top
of the plot. There is significant variability between the light curves which
is presumably the result of changes in spot coverage and location.  The
overall impression from Figs~\ref{fig:binary-fold} and
\ref{fig:multi-binary-fold} is of a fairly typical, late-type, main-sequence
eclipsing binary.

Our main interest in analysing the binary on its own is to establish as many
parameters as possible that can be held fixed in subsequent fits to the
dips. While a simultaneous fit of all parameters would be preferable, it turns
out that even for our best models the dips and the binary eclipses each point
towards very different values of some parameters. Examples are the ratio of
radii of the two stars of the close binary and its epoch of zero phase. These
are a few of several indications that our model of the dips must be
wrong. Therefore, starting from the premise that we understand the binary
better than the dips, we adopt the approach of letting the binary
data fix as many parameters as possible, before attempting to model the
dips. We used the same code as we later use in fitting the dips and will defer
a description of the methods until then.

\begin{figure}
\hspace*{\fill}
\centering
\includegraphics[angle=270,width=\columnwidth]{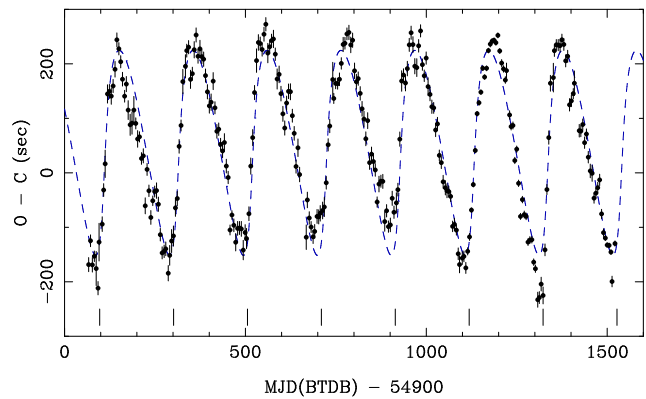}
\hspace*{\fill}

\caption{The observed times of the binary eclipses minus those calculated
  assuming a constant period show a large modulation due to the binary's orbit
  with a third object. The dashed line shows an eccentric orbit fit which was 
  established from a light-curve to all the data excluding those affected by 
  dips. The times plotted were
  calculated by holding all parameters fixed apart from the binary's zeropoint
  and fitting to 280 sub-sections of data, equally spaced from start to
  finish. The vertical lines at the bottom mark the mid-times of the dips
  (\kepler\ stopped observing \kic\ just before the last one indicated).
  For reference, light takes $499\,\s$ to travel $1\,$AU.
  \label{fig:ltt}}
\end{figure}

As part of fitting the binary's light curve, we had to allow for the timing
variations caused by its orbit within the triple, therefore the parameters
associated with the triple orbit are a natural by-product of the light curve
modelling.  In order to show the effect of the triple star's orbit, we
measured the epoch of the binary over small intervals (roughly five days) of
time throughout the \kepler\ observations. The results, along with the fit
derived from the binary light curve analysis are shown in Fig.~\ref{fig:ltt}.
The variations seen are highly significant. The dips occur close to the time
when the binary is closest to us, so that we see its eclipses arrive early.
There can be no doubt over \cite{2013ApJ...763...74L}'s conclusions that the
binary is in an eccentric orbit with another object and that the dips occur
when it passes in front of that object.

The parameters derived from the fits to the binary and its orbit within the
triple are listed in Table~\ref{tab:binary}.
\begin{table}
\begin{tabular}{lll}
Name & Value & Lee et al\\
\hline
$T_b$                    & $55632.530537(34)$   & $55632.53016(14)$ \\
$P_b$ [d]                & $0.2585073013(50)\,$ & $0.25850790(12)$ \\
$a_b \sin i_t$ [$\rsun$] & $99.38(70)$          & $97(9)$ \\
$e_t$                    & $0.6011(57)$         & $0.612(82)$ \\
$\omega_t$ [$^\circ$]    & $344.64(64)$          & $353(4)$ \\
\hline
$R_1/(a_1+a_2)$ & $0.2485(6)$   & $0.2041(45)$ \\
$R_2/(a_1+a_2)$ & $0.2268(8)$   & $0.3115(30)$ \\
$i_b$           & $89.66(24)^\circ$ & $85.32(66)^\circ$\\
$S_2/S_1$       & $1.155(2)$ & --- \\
$l_3$           & $0.97235(6)$ & $0.972(1)$\\
\hline
\end{tabular}

\caption{Parameters describing the inner binary and its orbit within the
triple derived from a fit to all the data excluding the triple star dips.
\label{tab:binary}}

\end{table}
We did not determine the triple ephemeris from the binary data as it is more
precisely pinned down by the dips (see section~\ref{sec:nagging}).
Where possible we
list the equivalent values from \cite{2013ApJ...763...74L}, although our two
models are not precisely the same since they used a more sophisticated model
accounting for tidal deformation, gravity darkening and star spots, which, for
reasons of compatibility with the triple star models to be described later, we
do not apply. We believe that in any case the degeneracy associated with the
spots limits the accuracy with which some parameters can be determined, as we
will detail shortly. It should be noted that we truncated the distribution of
$i_b$ at $90^\circ$ where it peaks. We distinguish between those parameters
which describe the binary's ephemeris and the orbit of its centre of mass
within the triple ($T_b$, $P_b$, $a_b$, $e_t$ and $\omega_t$) and those which
control the shape of the light curve, (the scaled radii $r_1 = R_1/(a_1+a_2)$
and $r_2 = R_2/(a_1+a_2)$, $i_b$, $S_2/S_1$ and $l_3$). The first set depend
upon timing information, and can only be fixed by a fit to the whole light
curve. When we come later to fit the dips, we will hold these fixed since any
individual set of dips contains little information to constrain them
(with the possible exception of $T_b$ and $P_b$ -- see
section~\ref{sec:nagging}). In
contrast the dips are highly sensitive to $r_1$, $r_2$, etc, which we will
call the shape parameters.

The shape parameters are not nearly as well constrained as the purely
statistical uncertainties listed in Table~\ref{tab:binary} suggest because
of distortion of the light curve, which like \cite{2013ApJ...763...74L}, we
put down to star spots. The effect of these is obvious outside eclipse in
Figs~\ref{fig:binary-fold} and \ref{fig:multi-binary-fold}. It can be seen
quantitatively in the poor agreement between our radius values and those of
\cite{2013ApJ...763...74L}, and even amongst different fits of our model. For
instance, we obtain an overall radius ratio (Table~\ref{tab:binary}) of
$R_1/R_2 = 1.10$, but when fitting to the short-cadence data alone we find
$R_1/R_2 = 0.88$ (Fig.~\ref{fig:binary-fold}).

We focus on the radius ratio in particular because it turns out that to fit
the dips we need much more extreme radius ratios than the values listed in
Table~\ref{tab:binary} suggest. In Fig.~\ref{fig:binary-fold} we show three
model light curves. The solid line model shows a best fit with all the shape
parameters allowed to vary. This led to $R_1/R_2 = 0.88$ and an almost edge-on
inclination. For the other two we force large radius ratios $R_1/R_2 = 2$ and
$R_1/R_2 = 4$ (which also forced much lower inclinations of $74^\circ$ and
$54^\circ$ respectively). Our object in plotting these is to show that even
large changes in the shape parameters, which are well outside the statistical
uncertainties of Table~\ref{tab:binary}, lead to relatively small changes in
the light curves which are comparable to the spot-induced variations. This
reflects a well-known degeneracy for partially-eclipsing, or almost
partially-eclipsing binaries. As a result when we fit the dips, we do not
restrict the shape parameters to the values listed in Table~\ref{tab:binary}
but merely apply the following constraint that ensures that eclipses of the
correct total width, $\Delta \phi = 0.155$ occur:
\begin{equation}
\left(r_1+r_2\right)^2 = \cos^2 i_b + \sin^2 i_b \sin^2 (\pi \Delta \phi) .
\end{equation}

Systematics are also visible in the light travel times of
Fig.~\ref{fig:ltt}. It again seems likely that these reflect the spot-induced
variability evident in Fig.~\ref{fig:multi-binary-fold}. Although we weighted
our fits towards the eclipses (since in our model the region out-of-eclipse
contains no useful information on the timing), we can still expect the
eclipses themselves to be affected by star spots. Given that the orbital
period is $22300\,$s, and the level of variability in the light curves, the
$20$ -- $50\,$s systematic deviations seen in Fig.~\ref{fig:ltt} are
understandable.  Having established the triple's orbit and the nature of the
close binary star's light curve, we now begin our analysis of the dips.

\subsection{The Dips}

\subsubsection{Overview of the Dip light curves}
\label{sec:overview}
\begin{figure}
\hspace*{\fill}
\centering
\includegraphics[angle=270,width=\columnwidth]{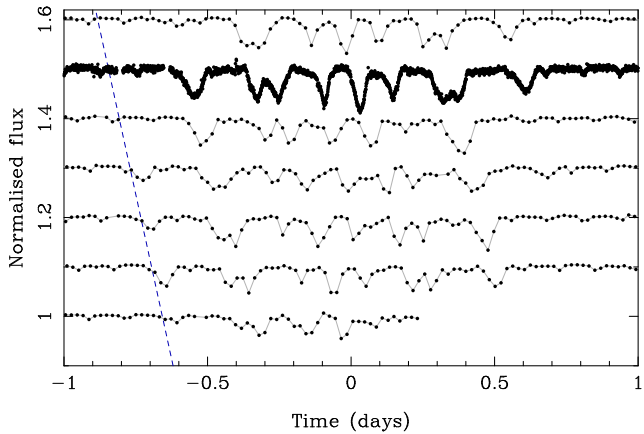}
\hspace*{\fill}

\caption{The seven consecutive dips in \kic\ observed by \kepler, with time
  running upwards.
Each has been aligned on the triple star ephemeris. From one set of dips to
the next, the same dips are seen to recur, but to arrive earlier relative to
the triple star ephemeris, as indicated by the dashed line which runs along
the left-edge of the earliest dips seen in the second, third, fourth, and
possibly also the first, set of dips. \label{fig:dips}}
\end{figure}

We start with a qualitative, model-independent assessment of the light curves
during the appearance of the dips. Fig.~\ref{fig:dips}
shows all seven of the occurrences of the dips observed by \kepler. No two
sets of dips are identical, but many bear strong similarities to each
other. For instance, counting from the bottom, the second, third and sixth
sets are very similar to each other, with each showing first a single dip,
followed by a closely-spaced double dip, then by three somewhat more widely
spaced dips, then another double dip and then a final single dip. The pattern
of dips of one set can be seen to arrive slightly earlier at the next set, as
indicated by the dashed line in fig.~\ref{fig:dips}. This can't happen while
maintaining a lock to the triple orbit without the dips evolving, and this can
seen as early arrivals fade away on the left-hand side of Fig.~\ref{fig:dips}
just as new dips grow in strength on the right-hand side.  (The situation is
reminiscent of the behaviour of wave crests in groups of ripples on the
surface of a pond.) The earliest any dip is seen is in the fourth event at
around $-0.72\,\d$; the latest is also the fourth event at around $+0.72\,\d$,
so the total duration of the dips exceeds $1.4\,\d$.

\begin{figure}
\hspace*{\fill}
\centering
\includegraphics[angle=270,width=\columnwidth]{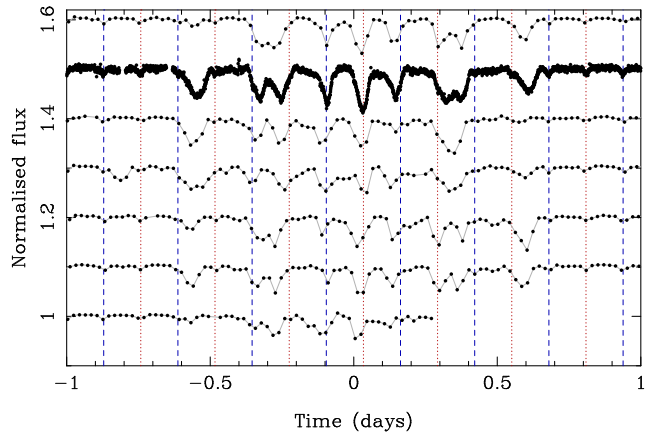}
\hspace*{\fill}

\caption{The same as Fig.~\protect\ref{fig:dips}, except now the times
have been adjusted to the nearest binary cycle. Dashed lines mark binary
phase~0; dotted lines mark binary phase 0.5. \label{fig:dips-binary}}
\end{figure}

In Fig.~\ref{fig:dips-binary}, 
we show the dips again but now with the times adjusted to the binary phase.
This plot shows some very interesting features. First, the dips are stable
relative to the binary phase. Second we see that the first dip in a given set
always occurs during the binary phase interval 0.0 to 0.5, while the last
always occurs in the interval 0.5 to 1.0. Moreover, no dip is seen in the half
cycle following the first dip, or in the half cycle preceding the last
dip. This is \emph{extremely odd}. In the previous section, we showed that the
two components of the binary have similar radii, so it should look very
similar at any two phases 0.5 cycles apart. However it seems instead that in
one configuration obscuration occurs while it does not in the other. It is as
if only one of the two stars in the binary occults the third object.

In contrast to the half-cycle following the first dip in a set, which seems to
be free of any obscuration, the half cycle following the double dip (covering
around $-0.22$ to $-0.1\,\d$cd in Fig.~\ref{fig:dips-binary}) is usually not
entirely clean, but shows some slight slopes. The half cycle preceding the
final double dip behaves similarly. These are perhaps small signs of the
presence of star~2, although the contrast between successive half cycles is
still much stronger than the binary model would suggest.

\subsubsection{Modelling the Dips}
\label{sec:dips}
In order to model the dips, we developed a model of a triple star involving
three limb-darkened spheres, in hierarchical Keplerian orbits, specified by
the parameters listed in Table~\ref{tab:param}. Our concern here is to capture
the main features of the data and we do not include effects of secondary
importance (for \kic\ at least) such as tidal distortion and gravity
darkening. This considerably speeds the computations, which are a limiting
factor in much of the modelling. We do not account for $N$-body corrections to
the Keplerian orbits because we expect these to be small given the $\sim
800$-fold ratio of the outer and inner orbital periods. To compute the flux
from each star, its circular projected face was split into a set of concentric
annuli of constant radial increment but variable surface brightness because of
limb darkening. The task then breaks down to working out how much of each
annulus is visible given the locations and sizes of the other two stars.  The
number of annuli determines the extent of numerical noise. We used 80 for each
star, verifying that the resultant numerical noise was less than the noise
level in the data. The only notable feature of the parameters chosen is our
choice of zeropoints, $T_b$ and $T_t$, which mark central times of the binary
star's primary eclipses on the one hand and the ``dips'' on the other, as
opposed to the more usual time of periastron passage. We made this choice
because the eclipse and event times are more-or-less directly fixed by the
data. This means that the resulting epochs are much less correlated with other
parameters than they would have been had we used the periastron times instead.
We tested the code by verifying that it correctly reproduced the light curve
of the triple system KOI-126 at the epoch closest to the epoch of the orbital
elements quoted by \cite{2011Sci...331..562C}.

\begin{figure*}
\hspace*{\fill}
\centering
\includegraphics[angle=270,width=0.7\textwidth]{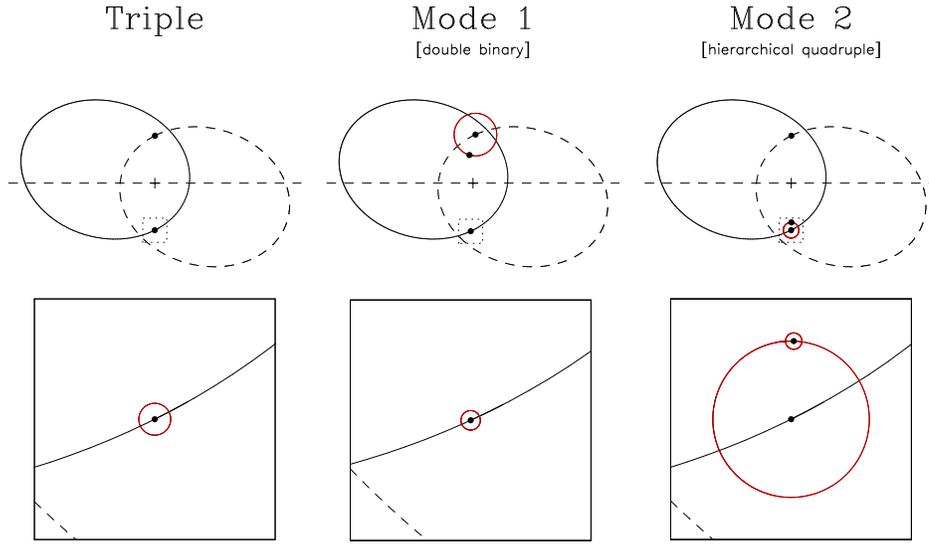}
\hspace*{\fill}

\caption{A face-on view of the three types of orbits (one triple, two
  quadruple) we consider for \kic, all drawn to scale with tilts removed so
  that there is no distortion by projection. The observer is assumed to lie in
  the plane of the figure and view the orbits edge-on from the bottom of the
  page. The horizontal dashed line represents the plane of the sky, with the
  '+' signs marking the system's centre of mass. The two ellipses show the
  centre of mass of the close binary in the left two panels (solid) and star~3
  in the left- and right-hand panels (dashed). The lower panels show factor 10
  magnified views of the regions delineated by small dotted squares centred
  near the close binary. In the central panel, the dashed ellipse shows the
  centre of mass of a second binary formed from star~3, the star that is
  eclipsed during the dips, and an extra star, ``star~4'', introduced in order
  to alter the dynamics for reasons explained in the text. The upper circle
  shows the orbit of star~3 around a fixed point on the dashed ellipse.  All
  orbits are traversed counter-clockwise in this figure, so that in the
  configuration shown, which matches the time of the dips, the relative
  transverse speed between star~3 and the close binary is slowed by stars~3's
  motion within its binary with star~4.  In the right-hand panel the
  solid-lined ellipse shows the track of the centre of mass of an inner
  triple made up of the close binary and ``star~4'', once more introduced to
  alter the dynamics of the system, but in a different configuration.  The
  close binary thus moves on a circle around this guiding centre, as indicated
  by the large circle in the right-hand magnified view. The small circles,
  which can only be seen in the magnified views, represent the orbit of 
  star~1 (again with its guiding centre held stationary
  at the time of the dips). Dots in each panel show the centre of
  mass of the close binary and star~3. Additionally, in the centre panel the
  dot on the dashed ellipse represents the centre of mass of the second
  binary, while in the right-hand panel the dot on the solid ellipse marks the
  centre of mass of the inner triple. The orbits of stars~2 and 4 are
  suppressed for clarity.
  \label{fig:orbits}}
\end{figure*}

Later on we will examine quadruple star models for \kic, making the geometry
harder to visualise. Therefore in Fig.~\ref{fig:orbits} 
we give schematic pictures of the three types of orbits that we will consider.
The observer is assumed to be in the plane of the figure, looking from below,
and the system is shown at the time of the dips.

We varied or fixed parameters according to whether the data significantly
constrained them. For instance, we set $\Omega_t = 270^\circ$ (ascending node
due West on the sky, in the usual direction of the $x$-axis) from the start
since there is no information upon the absolute orientation of the system. On
the other hand, the data strongly constrain the relative orientation the
binary and triple, so $\Omega_b$ was allowed to vary. Similarly, the limb
darkening coefficients were uniformly set equal to $0.5$ since they have a
relatively minor effect upon the light curves. In the case of the binary, the
nature of its light curve and its short period very much suggest that it has a
circular orbit, and so we fixed $e_b = 0$ and $\omega_b = 0$ for all models.
Finally when fitting to data, we scaled the fluxes to minimise $\chi^2$ as
this is a fast operation. This meant that one of the surface brightness
parameters could be fixed (since otherwise there would degeneracy between the
surface brightnesses and the scaling factor), leaving us with a maximum of
18 parameters that could be varied.

As previously explained, owing to clear differences between the parameter
space favoured by the binary light curve compared to the dips, we first fitted
those parameters which could be determined from the binary alone.  Thus we
fixed $T_b$ and $P_b$ which define the binary's ephemeris, and $a_b$, $e_t$
and $\omega_t$ which define the triple orbit, to the values listed in the top
section of Table~\ref{tab:binary}. We will see later that all lengths in the
system scale with the value of $a_b + a_3$, the semi-major axis of the triple,
and masses therefore scale as $(a_b+a_3)^3$. While $a_b$ is fixed by the light
travel times (given that $\sin i_t = 1$ to a good approximation), we have no
direct information upon $a_3$, although it can be estimated by seeking a
consistent set of masses and radii for the binary star, assuming it to be
composed of a pair of M dwarfs. This is because the binary light curve fixes
the radii scaled by the total separation, the masses scale as the total
separation cubed, while the mass and radius of low mass stars are nearly
linearly related \citep{2013AN....334....4T}. Assuming a precise linear
relation, $M/\msun = R/\rsun$, and starting from the values listed in
Table~\ref{tab:binary} leads to an estimate for $a_3 \approx 75 \,\rsun$.  We
therefore adopt a round number of similar magnitude, and henceforth will
assume that $a_3 = 100\,\rsun$. Where relevant later, we point out aspects
that depend upon the particular value chosen for $a_3$.

We carried out the fitting through a combination of standard minimisation
methods \citep{nelder-mead, powell} and (mainly) Markov Chain Monte Carlo
(MCMC) iteration. MCMC takes a Bayesian point of view whereby one constructs
models that are distributed with the posterior probability distribution of the
parameters, given the data. The posterior probability has prior probabilities
representing one's knowledge before any data are taken times the probability
of the data given the model. The latter is encapsulated by $\chi^2$ in our
case since we assume independent gaussian uncertainties on the data. The prior
provides a flexible way to impose physical constraints without requiring that
they hold precisely at all times during minimisation. The most important such
constraint comes from Kepler's laws. We implemented our model with a
combination of \texttt{C} and \texttt{Python}, and used the \texttt{emcee}
package \citep{2013PASP..125..306F} to manage the MCMC computations.

Kepler's third law applied to the triple orbit gives us the total system
mass in terms of the controlling scale factor, $a_3+a_b$:
\begin{equation}
G (m_1+m_2+m_3) = n_t^2 \left(a_3+a_b \right)^3, \label{eq:kepler1}
\end{equation}
where $n_t = 2\pi / P_t$. The centre-of-mass condition 
\begin{equation}
m_3 a_3 = (m_1+m_2) a_b, \label{eq:kepler2}
\end{equation}
then allows us to deduce the mass of each component of the triple. A second
application of Kepler's third law then fixes the total separation of the
binary
\begin{equation}
G (m_1+m_2) = n_b^2 \left(a_1+a_2 \right)^3, \label{eq:kepler3}
\end{equation}
where $n_b = 2\pi/P_b$. Thus the value of $a_1+a_2$ is fixed once
$a_3+a_b$ is fixed and cannot be allowed to vary independently of it. This we
ensured through the prior probability by demanding near-equality between
the value of $a_1+a_2$ computed as above starting from $a_3+a_b$ and the 
value derived from the models proposed during the MCMC process. Inequality was
punished through low prior probability. This method allows great flexibility
in terms of what is allowed to vary, whilst ensuring physical consistency. The
degree of equality demanded, which has some impact upon the MCMC efficiency,
could be tuned at will.

\begin{figure*}
\hspace*{\fill}
\centering
\includegraphics[angle=270,width=\columnwidth]{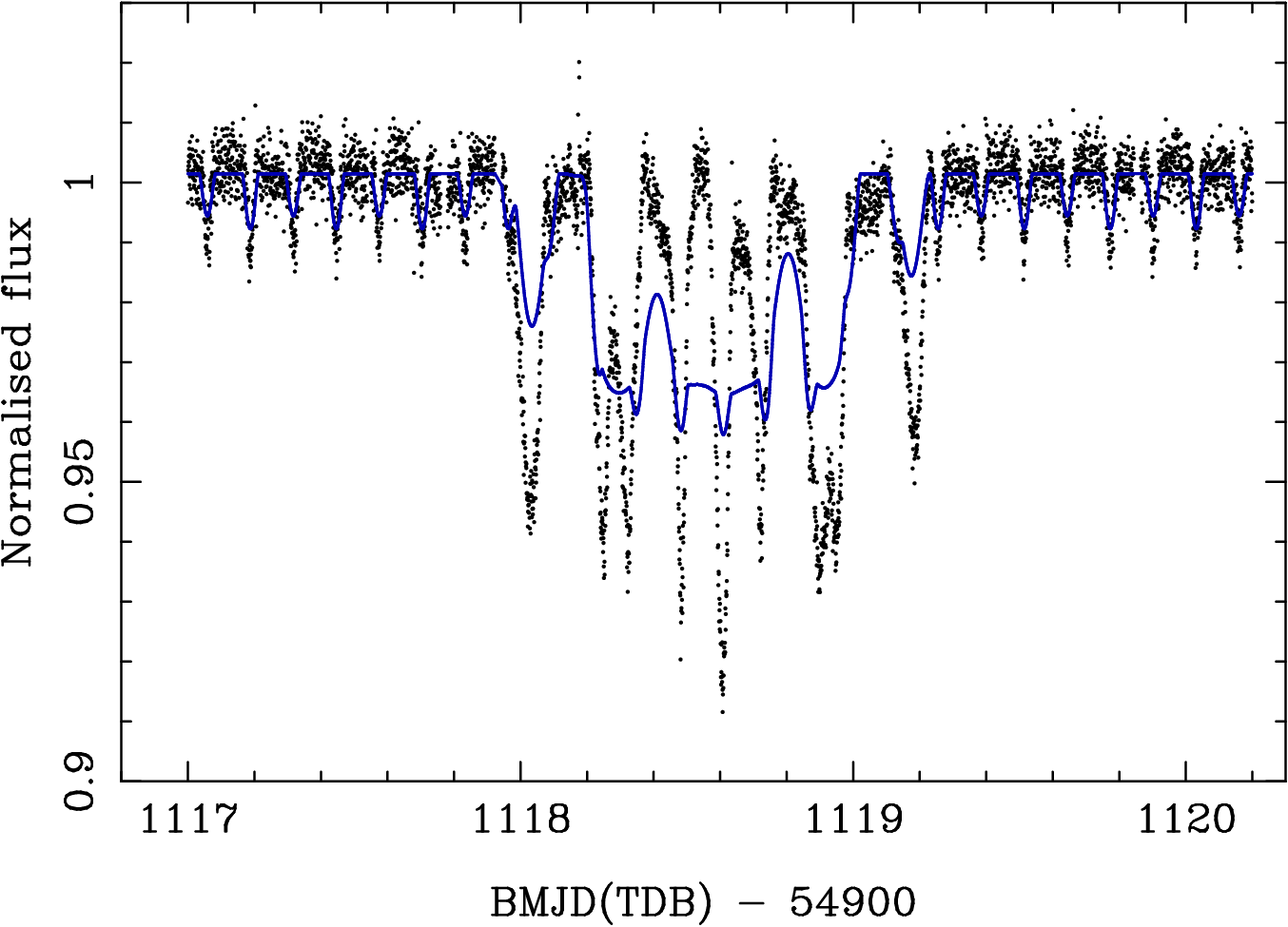}
\includegraphics[angle=270,width=\columnwidth]{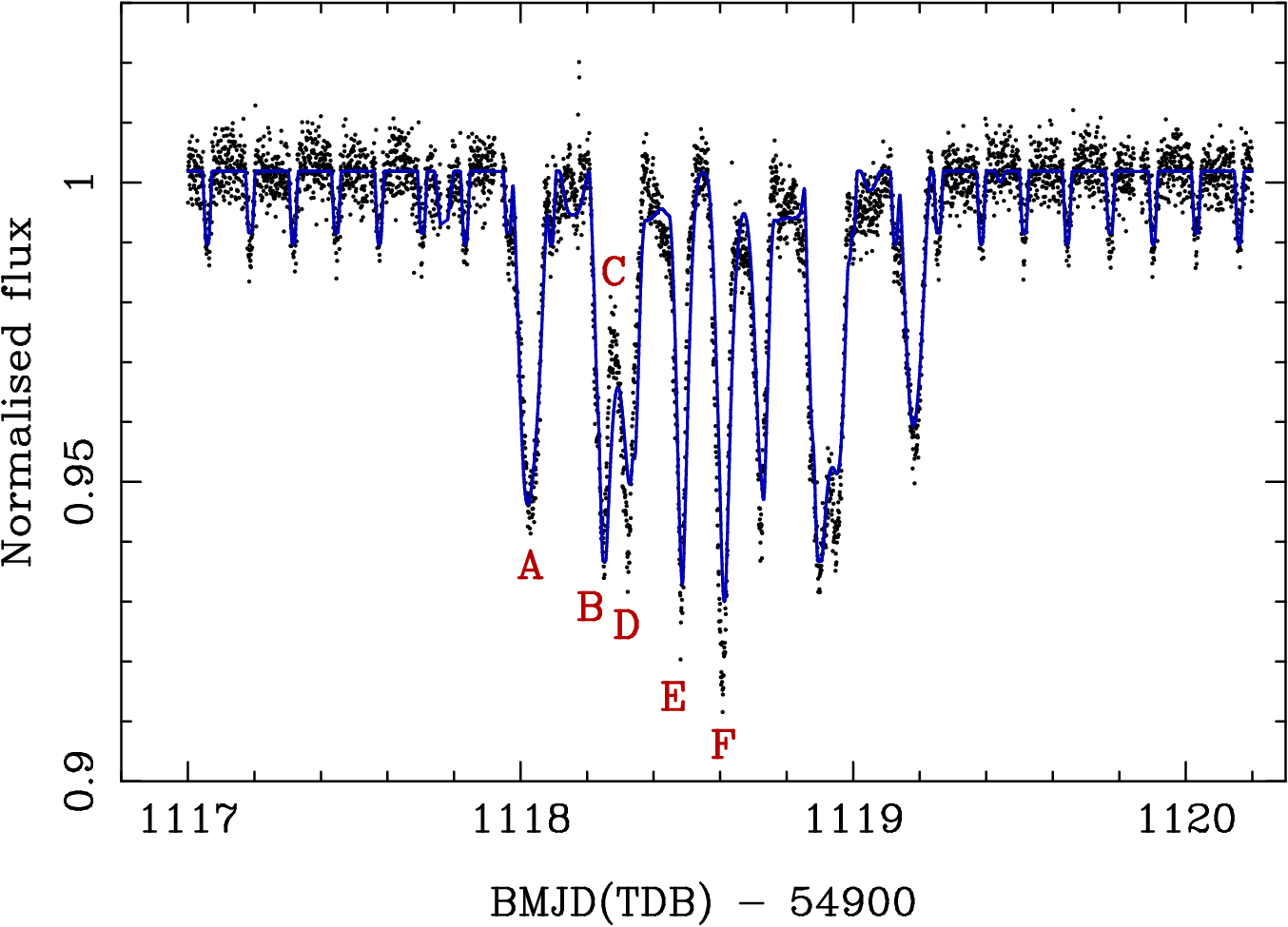}
\hspace*{\fill}
\caption{\emph{Left-panel:} The set of dips observed in short cadence
along with the best fitting triple star model of \kic\ with the system
parameters constrained to obey Kepler's Laws and to match the binary
eclipse width. \emph{Right-panel:} The same data and model, but with 
the Keplerian constraint removed. The letters label features
in the light curve for ease of description in the text.
\label{fig:triple}}
\end{figure*}

Fixing what parameters we could from the binary model, setting $a_3 =
100\,\rsun$, we went ahead and optimised the remaining 12 parameters which
were the stellar radii $R_1$, $R_2$ and $R_3$, the binary semi-major axes
$a_1$ and $a_2$, the orbital inclinations $i_b$ and $i_t$, the epoch of the
triple $T_t$, the surface brightness parameters $S_1$ and $S_2$ along with the
``third light'' $l_3$, and the orientation of the binary orbit, $\Omega_b$. We
applied Kepler's laws via the prior as just outlined, and the constraint upon
$r_1$, $r_2$ and $i_b$ to match the eclipse width that we described in
section~\ref{sec:binary}.  The best fit to the short cadence dips resulting
from this procedure is shown in the left-hand panel of Fig.~\ref{fig:triple}.
Clearly ``best'' is very much a relative term here as the fit is extremely
poor with $\chi^2 \approx 60000$ for $4589$ points. In particular the level of
modulation in the central part of the dips is much weaker in the model than
the data. This is because in the model $R_3 \approx 3.9\,\rsun$ is large
compared to $R_1 = 1.29\,\rsun$ and $a_1 = 1.85\,\rsun$. Once the binary
starts to cross star~3, one part of it is always in an occulting position.
Besides providing a poor fit, the parameters that lead to the fit shown in
Fig.~\ref{fig:triple} can be ruled out on astrophysical grounds. 
For instance star~1
ends up with almost zero mass ($10^{-5}\,\msun$) but a radius of $1.29\,\rsun$
(and thus overfills its Roche lobe).

\begin{figure}

\centering
\includegraphics[angle=270,width=0.9\columnwidth]{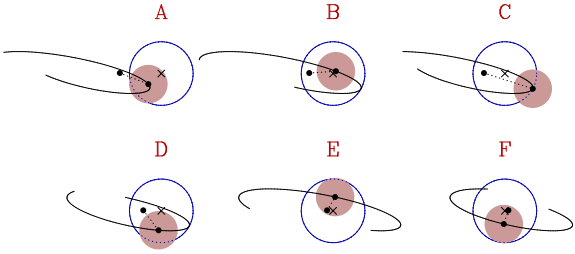}

\caption{The geometry of stars~1 (small and shaded) and 3 (large and outlined)
  at the times labelled in Fig.~\protect\ref{fig:triple}, according to the
  model shown in the right-hand panel of that figure. Star~1 is closer to the
  observer than star~3 at all times, and is blocking flux from it. The
  circular dots connected by dotted lines show the centres of mass of star~1
  and the binary at the time in question; the cross marks the centre of mass
  of star~3. The solid curves indicate the path of the centre of mass of
  star~1 over the range $-0.1$ to $+0.1\,$d relative to the particular time in
  question. They can be thought of as small sections of the projection of a
  squashed helix. The orbital inclination of the binary in this model $i_b >
  90^\circ$ so that the paths are travelled in a clockwise manner. For clarity
  we have shrunk the radius of star~1 (solid) by a factor of 2; this does not
  affect the times of minima / maxima which depend upon the impact parameter
  only. The precise radius ratio, which controls the depth of the dips,
  depends also upon the amount of ``third light'' and is therefore not well
  defined.
  \label{fig:geom}}
\end{figure}

Despite the poor fit, the model does show some similarities to the data
indicating that it contains elements of truth. A much better although still
imperfect fit ($\chi^2 = 13900$) is obtained if the Keplerian constraint is
removed, as shown in the right-hand panel of Fig.~\ref{fig:triple}. While
physically impossible, it is useful to understand how this model manages to
match the dips as well as it does. To facilitate our discussion of this
we label specific features of the lightcurve using the letters A to F as 
shown in the right-hand panel of Fig.~\ref{fig:triple}. Further we concentrate
upon how the largest star of the binary (star~1) transits star~3 because
star~2 in this model plays almost no part in the dips. This is forced 
by the peculiar absence of any obscuration between minima A and B that we
noted in section~\ref{sec:overview}. With this simplification, the geometry 
equivalent to the model of the right-hand panel of Fig.~\ref{fig:triple}
is shown schematically in Fig.~\ref{fig:geom}.
As time progresses, the centre of mass of the binary moves from left to right
in this figure. Minima in the light curve occur at the points of closest
approach (minimum impact parameter) between the centres of stars~1 and 3. The
first of these (with significant obscuration) occurs at time~A. Remembering
that the the binary executes its orbit rapidly compared to the advance of the
outer orbit, one can see that even if one altered the triple phase so that the
centre of mass was in a slightly different part of its path, the minimum flux
would always be located close to the binary quadrature phase, with the main
change being in the amount of overlap of the two stars at the time of
minimum. This explains the phenomenology we described when
discussing Fig.~\ref{fig:dips-binary}. For instance this is why the dips
appear to be locked to the binary rather than the triple phase, and why the
first dip is always seen near quadrature.

Times B, C and D span the first of the two double dips, with B and D marking
minima and C the intermediate maximum. These three times occur within the same
half orbit around quadrature. The single minimum of an A-like event is split
into two because while the centre of mass is on the left-hand side of star~3,
star~1 moves from the left-hand side to the right-hand side and then back
again, leading to two times of minimum impact parameter, with a point close to
quadrature where there is a local maximum in the impact parameter, giving a
maximum in the flux. Finally, when we have reached the central dips at E and
F, the centre of mass of the binary is close to half-way across star~3, meaning
that the points of minimum impact parameter are close to the times of binary
eclipse.

\begin{figure}

\centering
\includegraphics[angle=270,width=0.8\columnwidth]{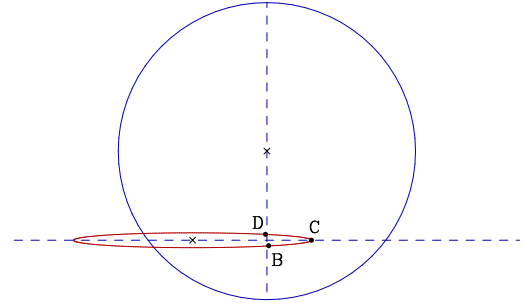}

\caption{The geometry that leads to the double dips. The ellipse shows the
projected  motion of the centre of star~1 around the binary's centre of mass 
(lower-left cross) which for simplicity we have assumed to be stationary. The
circle shows the outline of star~3, and the upper cross its centre. A double 
dip in the light curve occurs if star~1 crosses the diameter of star~3 marked with the
vertical dashed line from left to right and then back again. B and D mark
times of minimum flux while C marks a maximum. The binary inclination used
here is $i_b = 87^\circ$, and we have aligned the ascending nodes of the
binary and triple orbits. The horizontal dashed line indicates the path of the 
binary's centre of mass.
\label{fig:geom-ddip}}
\end{figure}

The explanation of the double dip is important for the discussion of the next
section, so in Fig.~\ref{fig:geom-ddip}
we show the essential feature of the geometry that leads to double dips in our
model. The key point is that the centre of star~1 must cross and re-cross the
diameter of star~3 that is approximately perpendicular to the binary star's
line of nodes. This line is vertical in Fig.~\ref{fig:geom-ddip} since we have set
$\Omega_b = 270^\circ$ to keep the figure as simple as possible. (Both
Figs~\ref{fig:geom} and \ref{fig:geom-ddip} are constructed relative to
star~3.) On egress from the dips, the same happens in reverse, i.e. the centre
of mass of the binary moves to the right of the vertical line in
Fig.~\ref{fig:geom-ddip}, while star~1 moves from right to left, crossing the
vertical dashed line, and then right once more, re-crossing the vertical line
for the last time.

Before we leave this section, it is worth noting that we searched hard for
hidden parts of parameter space that might resolve the poor fit shown in the
left of Fig.~\ref{fig:triple}, but to no avail. The fitting process was
well-behaved in that very different starting models would eventually reach the
same fit with the same parameters. Thus we are convinced that the left panel
of Fig.~\ref{fig:triple} is the best that a physically consistent triple star
model can do. For this reason we do not think that \kic\ can be as simple as a
triple. In the next section we back up this conclusion with a more analytical
treatment of the problem.

\subsection{The Dynamical--Geometrical Paradox}
Why does the triple model do such a poor job of explaining the dips? In the
previous section we found that we could only get somewhere near the data by
``relaxing'' Kepler's law, so that the binary separation could become larger
(than physically allowable). Obviously this is not acceptable, but the reason
for the problem perhaps contains pointers to solving it, and so it is of
interest to come to an analytical understanding of it. This also serves as a
reassurance that the problems of the previous section are real, and not simply
coding errors.

To begin, consider the geometry of Fig.~\ref{fig:geom-ddip}. Since the centre
of mass of the binary is slowly moving from left to right in this diagram, it
is almost inevitable that the circumstance shown will occur at some
point during a set of dips. The hard part is to ensure that there
are \emph{two} sets of double dips, as widely spaced in time as observed, with
the second set mirroring the first during the egress phase of the dips. From
Fig.~\ref{fig:geom-ddip}, ignoring possible tilts of the orbit relative to the
horizontal, this can only happen if the semi-major axis of star~1, $a_1$,
exceeds half the distance traversed by the centre of mass of the binary
between the occurrence of the double dips, which we define to take time
$\Delta t$. Quantitatively we find that we require
\begin{equation}
\frac{a_1}{a_3+a_b} > \frac{1+e_t \cos \nu_t}{\sqrt{1-e_t^2}}
\, \frac{n_t \Delta t}{2},
\end{equation}
where $\nu_t$ is the true anomaly of the triple orbit at the time of
the dips, which is related to the periastron angle via
$\nu_t = 3\pi/2 - \omega_t$. From our fits to the triple's orbit
(section~\ref{sec:binary}) we calculate the first term on the right-hand side
to be $= 1.449 \pm 0.018$.
Using the $4\sigma$ lower bound on this factor
and setting $\Delta t = 0.7\,\d$, as measured from the outermost minima of
the double dips, one then finds that
\begin{equation}
\frac{a_1}{a_3+a_b} > 0.0148. \label{eq:lower}
\end{equation}
Projection factors that arise if the binary's orbit is tilted with respect to
the triple or problems of exact timing only serve to increase this lower limit.

An alternative limit upon the same quantity is derivable from Kepler's third
law. Eliminating the masses from Eqs~\ref{eq:kepler1}, \ref{eq:kepler2} and
\ref{eq:kepler3} gives
\begin{equation}
\frac{(a_1+a_2)^3}{P_b^2} = \frac{(a_3+a_b)^2 a_3}{P_t^2},
\end{equation}
which can also be written as
\begin{equation}
\frac{a_1+a_2}{a_3+a_b} = \left(\frac{P_b}{P_t}\right)^{2/3}
\left(\frac{a_3}{a_3 + a_b}\right)^{1/3} .
\end{equation}
The final term on the right-hand side is $\le 1$, while $a_2 \ge 0$,
so we deduce that
\begin{equation}
\frac{a_1}{a_3+a_b} \le \frac{a_1+a_2}{a_3+a_b} \le
\left(\frac{P_b}{P_t}\right)^{2/3} = 0.0117.\label{eq:upper}
\end{equation}
The upper limit from Eq.~\ref{eq:upper} (dynamics) is lower than the lower
limit from Eq.~\ref{eq:lower} (geometry). This is the ``paradox'' of the
triple star model of \kic. There is no room to escape this conflict.  Indeed,
the equalities in Eq.~\ref{eq:upper} can only be met if $a_2 \ll a_1$ and $a_b
\ll a_3$. We don't expect either of these to hold true, so even the $0.0117$
is likely to be a significant over-estimate. Thus, just as we found with the
numerical models, in triple star models one can fit the data (roughly) or
Kepler's laws, but not both with the same model.

A less rigorous constraint, that points in the same direction, but does not
rely on our interpretation of the double-dips, can be deduced from the total
width of the events which was noted in section~\ref{sec:overview} to be
$\Delta w \approx 1.4\,\d$. The maximum width is obtained when all orbits
are aligned and star~1 is exactly at quadrature when it first contacts
star~3. In that case a similar argument to the constraint that led to
Eq.~\ref{eq:lower} implies that
\begin{equation}
\frac{a_1+R_1+R_3}{a_3+a_b} \ge 0.0296.
\end{equation}
It becomes difficult to satisfy this and Kepler's laws without making the
ratio of $R_1/R_3$ so small that one cannot match the dips. For example, if we
assume the two stars in the binary are M dwarfs following the typical
mass--radius relation of M stars, then $a_1$, $R_1$ and $a_3$ are determined,
and we find that the maximum depth of dips should be $\approx 1$\%\ compared to
the 8\%\ observed. This is easier to escape than the problem with Kepler's
laws, since an increase in $R_1$ implies a decrease in $R_3$ and thus a
significant increase in the maximum dip depth, $(R_1/R_3)^2$, but it is
suggestive of a similar problem, i.e. that the dips last too long for the
triple model to accommodate.

\subsection{Quadruple models}
Both the problems described in the previous section can be ascribed to a high
relative transverse speed between the centre of mass of the binary and star~3.
In a triple system this is a simple function of the orbital parameters and,
given the light travel time constraints, we have no freedom to alter it.  The
only plausible way we have thought of to change the transverse speed
significantly is by adding a fourth star. If such a star is coupled either to
star~3 to form a pair of binaries (which we label ``mode~1'', see
Fig.~\ref{fig:orbits}) or to the close binary to form a hierarchical quadruple
(mode~2, Fig.~\ref{fig:orbits}), the resultant orbital motion may slow the
relative speed between the centre of mass of the close binary and star~3,
thereby alleviating the problems of the previous section.

We implemented quadruple models along the lines of the triple star model, and
for the first time were able to find solutions that qualitatively agree with
the data while obeying Kepler's laws. The fits that result are visually
indistinguishable from the right-hand panel of Fig.~\ref{fig:triple} so we do
not show them. The quadruple model is no panacea, but it is the closest we
have come to explaining \kic. The new orbit comes at the cost of a fine-tuning
problem since we require its period to be close to an integer divisor of the
$204\,\d$ period (within $\sim 0.001$ of an exact integer ratio). This is needed
to ensure that the binary occults star~3 at the same part of the new orbit at
each set of dips so that the relative speed is always slowed down. If this
ratio is not perfect, then at some point in the future one can anticipate
considerable changes in the dips, which could change duration or disappear
altogether.

In addition to the fine tuning, which is at least not an impossibility,
some astrophysical problems remain. Table~\ref{tab:mr}
\begin{table}

\begin{tabular}{lccccccc}
            & $M_1$ & $R_1$ & $M_2$ & $R_2$ & $M_3$ & $R_3$ & $M_4$  \\
\hline
Triple      & 0.00  & 1.29  & 1.28  & 0.16  & 1.27  & 3.89 & ---   \\
Quad, mode 1& 0.50  & 0.60  & 0.78  & 0.16  & 0.36  & 0.51 & 0.91  \\
Quad, mode 2& 0.23  & 0.53  & 0.42  & 0.13  & 1.27  & 0.40 & 0.63  \\
\hline
\end{tabular}

\caption{Masses and radii in solar units 
of the three types of models for $a_3 = 100\,\rsun$ (the values are taken from
the best-fit models for each case). The radius of star~4 is
not listed as it is not constrained by any of the models.\label{tab:mr}}

\end{table}
lists the masses and radii of the three types of models we have considered.
The triple model listed is the one forced to obey Kepler's laws, so that we
can define masses consistently, which means that it corresponds to the poor
fit of the left-hand panel of Fig.~\ref{fig:triple}. When viewing this table,
the unknown value of $a_3$ which enters into the length that defines the
scale, $a_b + a_3$, should be recalled (section~\ref{sec:dips}), so that all
radii are subject to an unknown scale factor $s$ relative to the values
listed, and all masses to its cube, $s^3$.  Since $a_3 > 0$ and we used $a_b =
99.4\,\rsun$ and $a_3 = 100\,\rsun$ in the table, then we can assert that $s >
0.5$, with $s = 1$ for the values listed in the table.  Of the three models,
the triple model can be ruled out astrophysically as well as from its poor fit
to the data, as we noted earlier, since it implies almost zero mass for star~1
($10^{-5}\,\msun$ to be exact). This comes about from a vain effort to make
$a_1$ as large as possible, at the expense of $a_2$ to match Kepler's
laws. The small size of star~2 is probably the thorniest issue for the
quadruple models. For the mode~1 model, the large mass of star~4 is also a
potential problem, since it might end up dominating the light from the system,
even though it does not participate in creating the variations seen. However,
both quadratic models appear to need significant ($\sim 70$\%) ``third light''
contributions in addition to the light contributed by stars~1, 2 and 3, so
this may not be an impossibility. Otherwise, apart from star~2, the mode~1
mass--radius values look slightly preferable to those of mode~2, although
there is not a great deal to choose between them.  A final point against
mode~2 is that we expect an $\approx 20\,$ second semi-amplitude variations in
the light-travel times on the intermediate period (which in our fits lies in
the range $5$ to $20\,\d$). We searched for such a signal in the light travel
times of Fig.~\ref{fig:ltt} and were sensitive to semi-amplitudes of $10\,\s$
or more, but did not find anything.

In summary, the quadruple model is shaky, but it does at least not violate
basic physics in the same fashion as the triple model when getting close to
the data. In implementing the quadruple model, we continued to use hierarchical
Keplerian two-body orbits, but it is now much less clear that Newtonian
effects can be still neglected. Indeed if the quadruple model is correct,
Newtonian perturbations will need consideration to establish the dynamical
stability of the system. However, we did not try to add them because the number of
parameters of the quadruple model outstrips our ability to constrain them, and
because the poor fit to just the short cadence set of dips alone (during which
there would be no significant dynamical evolution) suggests that our model is
currently lacking important ingredients beyond $N$-body perturbations.

\subsection{Nagging problems}
\label{sec:nagging}
During our attempts to fit \kic\ two other peculiar problems surfaced that we
never resolved. The elusive second component of the binary is one of these. In
all the models that provide a reasonable fit to the dips, star~2 is 2.5 to 4
times smaller than star~1. We mentioned this as the chief drawback of the
quadruple models in the previous section, because there star~2's small radius
is out of kilter with its mass. Such unequal ratios are also not favoured by
the binary light curve, although in section~\ref{sec:binary} we argued that
star spots made the true ratio uncertain. However, there can be no doubt that
the two eclipses in the binary light curve have a similar depth, implying a
similar surface brightness for each component of the binary. It is then very
hard to understand how these two stars, which are most likely unevolved, low
mass main-sequence stars, can differ so much in radius.

\begin{figure}
\hspace*{\fill}
\centering
\includegraphics[angle=270,width=\columnwidth]{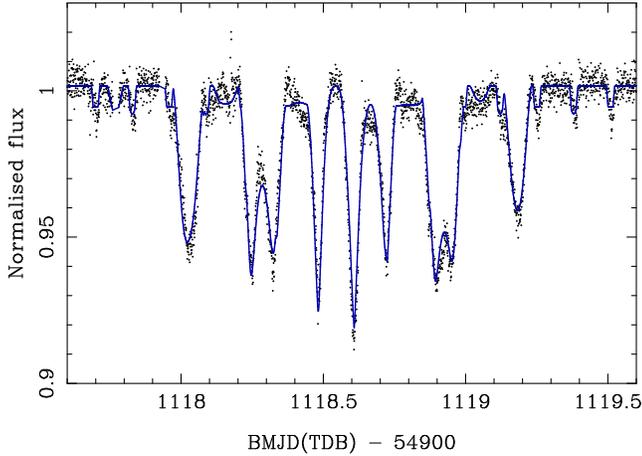}

\caption{The best mode~1 quadruple model fit to the short cadence dips when
the binary epoch $T_b$ is allowed to vary. The $\chi^2$ drops from $13600$ 
to $\approx 9000$, but the model binary eclipses arrive too early 
compared to the observed ones by $\approx 455\,\s$, a highly significant
shift. \label{fig:t0free}}
\end{figure}

\begin{figure}

\centering
\includegraphics[angle=270,width=\columnwidth]{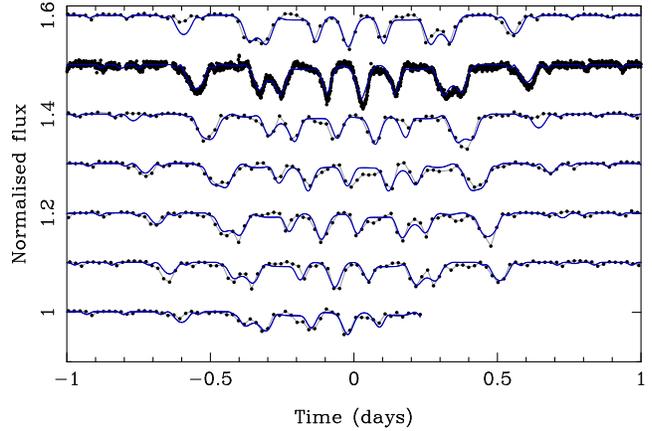}

\caption{All seven sets of dips together with a mode~1 quadruple model,
  accounting for finite exposure smearing for the long-cadence data.  The
  binary ephemeris was allowed to vary causing significant mismatches between
  the model and observed eclipses away from the dips which seem worst for the
  first set of dips. There is a qualitative match to the seven sets of dips,
  but also many significant discrepancies.
\label{fig:dips-fit}}
\end{figure}

Another very puzzling problem concerns the binary ephemeris.  None of the
models described so far provide a particularly good fit to the data, with,
at best, $\chi^2$ values around $13600$ for the 4589 points covering the short
cadence dips. A very odd feature is that this can be greatly improved,
with $\chi^2$ decreasing to $\approx 9000$, simply by letting the close binary
epoch $T_b$ vary (Fig.~\ref{fig:t0free}).
This comes at the significant cost of a mis-alignment between the
observed and model binary eclipses.  We examined this further by freeing up 
both $T_b$ and $P_b$ to fit all seven sets of dips simultaneously.
Fig.~\ref{fig:dips-fit}
shows the result. This fit returns a significantly longer binary period $P_b$
than obtained from fitting the binary only data (Table~\ref{tab:binary}), with
$\Delta P = (3.05 \pm 0.03) \times 10^{-6}\,$d, corresponding to $1250\,\s$
difference over the time between the first and last dips. This suggests that
the offset to the binary ephemeris that best fits the dips changes with
time. We have no solution to this curious problem which is perhaps another
clue to finding an improved model for \kic.

We used the triple model without Keplerian constraint to fit all seven sets of
dips in order to establish the ephemeris for the outer orbit which we used when
fitting the binary-only data in section~\ref{sec:binary}. We found
\begin{equation}
\mbox{BMJD} =  56018.5661(18) + 204.2723(9) E,
\end{equation}
where $E$ is an integer, giving the mid-point of the dips, with $E = 0$
coinciding with the set of dips observed in short cadence. We used the triple
model because the quadruple model introduces an extra degree of freedom which
renders these values much more uncertain, but the possibility of such
uncertainty should be borne in mind. Testing for this is a strong reason to
attempt ground-based observations of the dips. In using this ephemeris 
to predict future occurrences of dips, note the use of modified Julian days 
($\mbox{MJD} = \mbox{JD} - 2400000.5$).

\subsection{A spectral type for \kic}

\begin{figure}

\centering
\includegraphics[angle=270,width=\columnwidth]{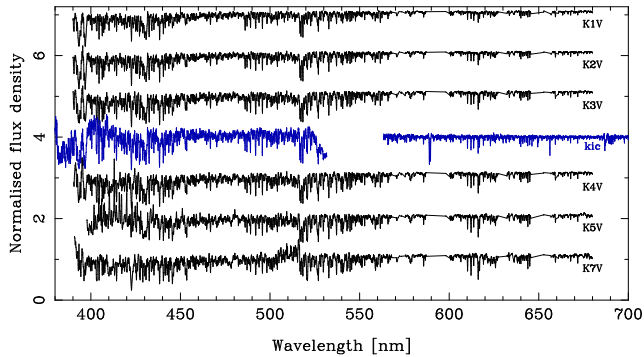}

\caption{The spectrum of \kic\ (centre) observed with the ISIS spectrograph on
  the William Herschel Telescope on May 21, 2012, compared to the spectra of
  main-sequence K stars \protect\cite{2001A&A...369.1048P}. Some mis-matches are caused by missing data in the templates (at
  H$\alpha$ and NaI~D), and by the dichroic cut between the blue and red arms
  of ISIS, but otherwise inspection of temperature-sensitive features
  suggests either a K3 or K4 type for \kic. We normalised the spectra by
  spline division and blurred and rebinned the template spectra to match
  the resolution of the ISIS data. \label{fig:spectral-type}}
\end{figure}

We obtained spectra of \kic\ on the night of May 21, 2012. The spectra were
taken during service time, in two sets, three hours apart. The spectrum did
not change significantly in this time, and so in Fig.~\ref{fig:spectral-type}
we present the normalised average of the two spectra. We compare these to
stars of known spectral type taken from the ELODIE.3.1 library of stellar
spectra which contains 1962 spectra of 1388 stars taken with the ELODIE
spectrograph at the Observatoire de Haute-Provence 193cm telescope in the
wavelength range $390$ to $680\,$nm \cite{2001A&A...369.1048P}. We used the $R
= 42000$ spectra from this library, which we blurred and re-binned to match
our data. Fig.~\ref{fig:spectral-type} shows that \kic\ is dominated by light
from a K3 or K4 star. In comparing this with the masses listed in
Table~\ref{tab:mr}, the unknown scale factor should be recalled.

\section{Discussion}
What at first appears to be a fortuitously aligned, but essentially
straightforward, triple system, raises a host of problems when one tries to
fit the dips that occur when the binary transits its tertiary companion. In
some despair, we turned to quadruple star models for the system, although
our faith in them is limited as they feel contrived  -- epicycles spring to
mind. Quadruple models have their own set of problems, although they
are less show-stopping than those which afflict the triple model.

Further observations are essential to guide future modelling of \kic.
Spectroscopy over several days could test the binary nature of star~3.
Spectroscopy over the $204\,$d cycle can provide a measurement of $a_3$ and
thus the total system mass. We expect radial velocity variations of several
tens of kilometres per second, and the spectrum (Fig.~\ref{fig:spectral-type})
has plenty of sharp line features which should allow precise radial
velocities.  Spectroscopy at long wavelengths might reveal the close
binary. The latter contributes a minimum of $\sim 3$\%\ of the light in the
\kepler\ bandpass, but this is poorly constrained and could be larger.  Even
at the minimum contribution, if the stars in the binary are M stars, then
given the mid-K star of Fig.~\ref{fig:spectral-type}, we can expect a
significantly higher contribution from the binary in the $I$- and $J$-bands.
Further monitoring of the dips, possible from the ground given the 8\%\
maximum depth, will also be of value to see whether they evolve significantly
with time. If the system is truly a quadruple star, then the simultaneous
short, long and intermediate orbital periods, together with the binary
eclipses and dips and highly eccentric outer orbit, suggest that it may be of
interest for dynamical studies, and significant evolution of the dips can be
expected.

\section{Conclusions}
\kic\ is an apparent triple star containing a close binary in orbit with
another object. Its orientation is such that the close binary passes in front
of its companion causing the appearance of a series of dips in the light curve
that last for a little over one day and recur on a period of $204\,\d$.  The
light curve of the binary is consistent with a pair of fairly well detached
and similar low-mass M dwarfs. While we expected the system to be
straightforward to understand, we were entirely unable to model it as a triple
star. Under triple models, the dips can only be modelled with separations of
the binary which violate Kepler's laws. Quadruple star models, involving
either two binaries in orbit around each other, or a binary orbited by another
star with another star orbiting the three of them, can match the data without
straining Kepler's laws, but require a very near integral ratio between the
$204\,\d$ period that the dips recur on and the period of the additional
orbit. There are moreover significant remaining mismatches between the model
and data even with the extra freedom provided by quadruple systems, and the
derived stellar parameters do not seem astrophysically plausible. \kic\ thus
defies easy explanation.  We urge further observations to uncover the true
nature of this remarkable object.

\section*{Acknowledgments}
TRM was supported under a grant from the UK's Science and
Technology Facilities Council (STFC), ST/L000733/1.
PJC acknowledges support provided by an Early Career Fellowship
from the Institute of Advanced Study, University of Warwick.
\kepler\ is NASA's tenth Discovery mission with funding
provided by NASA's Science Mission Directorate.
The WHT is operated on the island of La Palma by the Isaac Newton Group in the
Spanish Observatorio del Roque de los Muchachos of the Instituto de
Astrof\'{i}sica de Canarias.
\bibliography{refs}{}
\bibliographystyle{mn_new}

\label{lastpage}

\end{document}